\documentclass{JHEP3}

\usepackage{amsmath}
\usepackage{bm}
\usepackage{amssymb}
\usepackage{epic}
\usepackage{mathrsfs}
\usepackage{concmath, palatino}

\newcommand{\fldZ}{\mathcal{Z}}

\newcommand{\fldD}{\mathcal{D}}

\newcommand{\alg}[1]{\mathfrak{#1}}

\newcommand{\gym}{g\indups{YM}}
\newcommand{\indups}[1]{_{\mathrm{\scriptscriptstyle #1}}}
\newcommand{\gaba}{\gamma\inddowns{ABA}}
\newcommand{\inddowns}[1]{^{\mathrm{\scriptscriptstyle #1}}}

\newcommand{\ex}{\Upsilon}
\newcommand{\sca}{S_{0}}

\newcommand{\intd}{Y_{Q}}

\newcommand{\Baxter}{P^{(0)}_M}
\newcommand{\DBaxter}{P^{(0)'}_M}
\newcommand{\Baxtersub}{P^{(2)}_M}

\newcommand{\p}{\partial}
\newcommand{\Tr}{{\rm Tr \,}}
\newcommand{\Op}{\mathcal{O}}
\newcommand{\superN}{\mathcal{N}}
\def\MHSB{\hat\Omega}
\def\HSB{\Omega}
\def\HBS{\mathbb S}
\def\HS{S}
\def\reciP{\mathcal{P}}
\def\reciT{\mathcal{T}}

\newcommand{\beq}{\begin{equation}}
\newcommand{\eeq}{\end{equation}}
\newcommand{\beqa}{\begin{eqnarray}}
\newcommand{\eeqa}{\end{eqnarray}}

\newcommand{\sfrac}[2]{{\textstyle\frac{#1}{#2}}}

\newcommand{\M}{M}

\title{Six-loop anomalous dimension of twist-three operators in $\superN=4$ SYM}

\author{V.~N.~Velizhanin\\
Theoretical Physics Department\\
Petersburg Nuclear Physics Institute\\
Orlova Roscha, Gatchina\\
188300 St.~Petersburg, Russia\\
E-mail: \email{velizh@thd.pnpi.spb.ru}
}

\abstract{
The result for the six-loop anomalous dimension of twist-three operators in the planar $\superN=4$ SYM theory is presented. The calculations were performed along the paper arXiv:0912.1624. This result provides a new data for testing the proposed spectral equations for planar AdS/CFT correspondence.}

\def\zt{\zeta(3)}
\def\zf{\zeta(5)}
\def\zs{\zeta(7)}
\def\zn{\zeta(9)}

\begin{document}

\section{Introduction}
\label{sec:intro}

In the last two years the several higher-loop calculations of the different finite-length operators in $\superN=4$ supersymmetric Yang-Mills (SYM) theory were
performed~\cite{Fiamberti:2007rj}-\cite{Lukowski:2009ce}. 
These calculations usually consist of two part.
The first part comes from the Bethe Ansatz for the integrable spin chain, which was found first~\cite{Minahan:2002ve}\footnote{Earlier, the similar integrability was opened
in quantum chromodynamics in the Regge limit~\cite{Lipatov:1993yb,Lipatov:1994xy,Faddeev:1994zg} and for some of operators~\cite{Braun:1998id,Belitsky:2005bu}.}
in the leading order for the simplest BMN-operators~\cite{Berenstein:2002jq}.
Generalization to the higher orders together with the investigations of the integrable structures from the superstring theory side, started in Ref.~\cite{Bena:2003wd},
allowed to formulate all-loop Asymptotic Bethe Ansatz (ABA)~\cite{Beisert:2003tq}-\cite{Beisert:2006ez}.
However, as was shown explicitly in Ref.\cite{Kotikov:2007cy} by comparison with the predictions from the Balitsky-Fadin-Kuraev-Lipatov (BFKL)
equation~\cite{Lipatov:1976zz,Kuraev:1977fs,Balitsky:1978ic} the ABA gives incomplete result starting from the four loops for the twist-2 operators and
the wrapping corrections should be take into account.
Thus, the second part is the calculation of the wrapping corrections for the finite-length operators.
The first such calculation was performed perturbatively for the four-loop Konishi operator~\cite{Fiamberti:2007rj,Fiamberti:2008sh}
by taking only relevant Feynman diagrams, following Ref.\cite{Sieg:2005kd}.
Then, the wrapping correction for the four-loop Konishi was calculated using
the generalization of the L\"uscher formulae~\cite{Luscher:1985dn,Luscher:1986pf} for the finite-size corrections~\cite{Bajnok:2008bm}.

The result of both methods was confirmed with the full direct perturbative calculations without any assumptions at all (even without ABA)
by evaluation more then 130.000 diagrams~\cite{Velizhanin:2008jd}.
This result is an excellent perturbative test both for the ABA with the dressing factor and for the different methods of calculations of the wrapping corrections.

Then, the general form of the four-loop anomalous dimension for the twist-2 operators was obtained and the general structure for the leading finite-size
correction was understudied~\cite{Bajnok:2008qj}\footnote{The leading transcendental contribution was obtained earlier by direct perturbative calculation in~\cite{Velizhanin:2008pc}.}.
This result was used to calculate the general form of the five-loop anomalous dimension of twist-3 operators~\cite{Beccaria:2009eq},
which was confirmed by perturbative calculation in the most simple case~\cite{Fiamberti:2009jw}.

At the same time, few models for the complete spectral equations were proposed
in the form of the Y-system~\cite{Gromov:2009tv,Gromov:2009bc} and the Thermodynamic Bethe Ansatz (TBA) equations for the ground state~\cite{Arutyunov:2009zu}-\cite{Arutyunov:2009ax}.
They give the same results, which were obtained earlier with L\"uscher's formalism.
Nevertheless, it is necessary to check these all-loop proposals at the higher orders of perturbation theory.
New test for the proposed equations was obtained after the calculation of the five-loop Konishi with L\"uscher formalism~\cite{Bajnok:2009vm}.
In our previous paper with Tomasz \L ukowski and Adam Rej~\cite{Lukowski:2009ce} we have
generalized this result for the twist-2 operators with the arbitrary number of the covariant derivatives.
The importance of the obtained result is related with the exceptional properties of the twist-2 operators, namely, there is an excellent test for this result, originating from the BFKL equation, which gives prediction at any loop order for the analytical continuation of the complete anomalous dimension.
We have obtained full agreement with both the leading order and the next-to-leading
order~\cite{Fadin:1998py,Kotikov:2000pm} BFKL predictions. The most important consequence of the correctness of obtained result is the general structure of the modification of ABA due to the wrapping corrections.
Recently, using the same conditions for the ABA modification the result for the five-loop
Konishi was computed from the TBA equations~\cite{Arutyunov:2010gb,Balog:2010xa}.
Having in hands all necessary formulae and well working programs for the calculations of the first L\"uscher correction at the next-to-leading order for the twist-2 operators it is naturally to extend them to the analogies calculations for the twist-3 operators, as was done earlier at the leading order in Ref.\cite{Beccaria:2009eq}.

So, in this paper we generalize our previous calculations to the twist-3 operators at six-loop order.
First, we calculate the part of anomalous dimension, which comes from ABA.
Then, we write down explicitly all equations that need to be modified at transition from the twist-2 case to the twist-3 case.
After that, we present the result of finite-size  corrections with some details of calculations.
In the end we will check obtained results with known constraints.

\section{The six-loop anomalous dimension from Bethe ansatz}
\label{sec:fiveloop}

Twist-three operators belong to the $\mathfrak{sl}(2)$ sub-sector of the full theory.
They can be represented by an insertion of $M$ covariant derivatives $\fldD$ into the protected half-BPS state
\begin{equation}
\label{twistthree}
\Tr\left(\fldD^{m_1}\,\fldZ\,\fldD^{m_2}\,\fldZ\,\fldD^{m_3}\,\fldZ\right)+\ldots\,,\qquad m_1+m_2+m_3=M\,.
\end{equation}
In the spin chain picture such operators are identified with the states of the non-compact $\alg{sl}(2)$ spin $=-\sfrac{1}{2}$ length-three Heisenberg magnet with $\M$ excitations.
The anomalous dimension $\gamma(g)$
\begin{equation}
\label{dimension}
\gamma(g)=\sum_{\ell=1}^\infty  \gamma^{}_{2\ell}\,\, g^{2\ell}\,,\qquad
g^2=\frac{N\, \gym^2}{16\,\pi^2}\, ,
\end{equation}
may be determined to the four-loop order $\Op(g^8)$ with the help of the ABA~\cite{Staudacher:2004tk,Beisert:2005fw,Beisert:2006ez} and can be found in Refs.\cite{Kotikov:2007cy,Beccaria:2007cn}.
Starting from the five-loop order ABA gives the incomplete result due to wrapping corrections, which we will discuss in the next section.
So, the anomalous dimension can be split into the ABA part and the contribution of the wrapping interactions
\begin{equation}
\label{totalad}
\gamma(M)=\gamma^{\textrm{ABA}}(M)+\Delta_w(M)\,.
\end{equation}

The long-range asymptotic Bethe equations for the $\mathfrak{sl}(2)$ operators can be found directly from the full set of the asymptotic Bethe equations proposed in \cite{Beisert:2005fw,Beisert:2006ez}
\begin{equation}
\label{sl2eq}
\left(\frac{x^+_k}{x^-_k}\right)^L=\prod_{\substack{j=1\\j \neq k}}^\M
\frac{x_k^--x_j^+}{x_k^+-x_j^-}\,
\frac{1-g^2/x_k^+x_j^-}{1-g^2/x_k^-x_j^+}\,
\exp\left(2\,i\,\theta(u_k,u_j)\right),
\qquad
\prod_{k=1}^M \frac{x^+_k}{x^-_k}=1\, .
\end{equation}
There are $M$ equations for $k=1,\ldots,\M$ which need to be solved for the Bethe roots
$u_k$. The variables $x^{\pm}_k$ are related to $u_k$ through
\begin{equation}\label{definition x}
x_k^{\pm}=x(u_k^\pm)\, ,
\qquad
u^\pm=u\pm\tfrac{i}{2}\, ,
\qquad
x(u)=\frac{u}{2}\left(1+\sqrt{1-4\,\frac{g^2}{u^2}}\right).
\end{equation}
The function $\theta(u,v)$ is the dressing phase and has been conjectured in~\cite{Beisert:2006ez}. To the sixth order in perturbation theory it has the following form
\begin{eqnarray}
\label{4loopphase}
\theta(u_k,u_j) &=&\left(4\,
\zeta(3)\,g^6  -40\, \zeta(5) g^8 + 420\, \zeta(7) g^{10} \right) \big(q_2(u_k)\,q_3(u_j)-q_3(u_k)\,q_2(u_j)\big)\nonumber\\[2mm]
&&-\ 8\, \zeta(5) g^{10} \big(q_2(u_k)\,q_5(u_j)-q_5(u_k)\,q_2(u_j)\big)\nonumber\\[2mm]
&&+\ 24\, \zeta(5) g^{10} \big(q_3(u_k)\,q_4(u_j)-q_4(u_k)\,q_3(u_j)\big)
+\Op(g^{12})\, ,
\end{eqnarray}
where $q_r(u)$ are the eigenvalues of the conserved magnon charges,  see \cite{Beisert:2005fw}. The asymptotic all-loop anomalous dimension is given by
\begin{equation}
\label{dim}
\gaba(g)=2\, g^2\, \sum^{\M}_{k=1}
\left(\frac{i}{x^{+}_k}-\frac{i}{x^{-}_k}\right) .
\end{equation}

At one loop the Bethe roots $u_k$ are given by zeros of the Wilson polynomial \cite{Kotikov:2007cy,Beccaria:2007cn}
\begin{equation}
\label{eq:Wilsonpolynomial}
    P_M(u)={}_4 F_3\left(\left. \begin{array}{c}
    -\frac{M}{2}, \ \frac{M}{2}+1,\ \frac{1}{2}+iu ,\ \frac{1}{2}-iu\\[3mm]
    1,\  1 ,\ 1 \end{array}
    \right| 1\right) \,.
\end{equation}
Closed expressions for the corrections to the Bethe roots to the three-loop order have been obtained  in~\cite{Kotikov:2008pv} from the Baxter approach~\cite{Belitsky:2006wg}, which were also extended to the  higher-loop orders~\cite{Belitsky:2009mu}.

In order to obtain closed expressions for the anomalous dimension we solve
Eq.(\ref{sl2eq}) perturbatively for fixed values of the spin $M$ and match the coefficients
in an appropriate ansatz, which assumes the maximal transcendentality
principle~\cite{Kotikov:2002ab}\footnote{The hypothesis about maximal transcendentality principle~\cite{Kotikov:2002ab} was confirmed by direct perturbative calculations at the two-loop order~\cite{Kotikov:2003fb} and then successfully applied at the three-loop order~\cite{Kotikov:2004ss}, when corresponding results were obtained in QCD~\cite{Moch:2004pa,Vogt:2004mw}.}.
The basis for the ansatz is formed from the harmonic sums, which are defined by the following recurrent procedure (see \cite{Vermaseren:1998uu})
\begin{eqnarray} \label{vhs1}
S_a (M)&=&\sum^{M}_{j=1} \frac{(\mbox{sgn}(a))^{j}}{j^{\vert a\vert}}\, , \\[3mm]
S_{a_1,\ldots,a_n}(M)&=&\sum^{M}_{j=1} \frac{(\mbox{sgn}(a_1))^{j}}{j^{\vert a_1\vert}}
\,S_{a_2,\ldots,a_n}(j)\, .\label{vhs}
\end{eqnarray}
In the twist-three case the harmonic sums entering into basis should have all positive indices and argument $M/2$ (see Refs.\cite{Kotikov:2007cy,Beccaria:2007cn}).
The number of such sums at the $k$-loop order for the transcendentality level $2k-1$
is equal to $4^{k-1}$.
Up to four-loop order we need $4^3=64$ harmonic sums and the result for the anomalous dimension can be easily found (see Refs.\cite{Kotikov:2007cy,Beccaria:2007cn}).
However, the number of sums grows rapidly and already at five loops we have 256 harmonic sums in the basis, while at six loops the basis will contain 1024 harmonic sums.
Fortunately, the reciprocity~\cite{Dokshitzer:2005bf,Dokshitzer:2006nm} enters into the game. The presence of some structures in the anomalous dimension of twist-2 operators in $\superN=4$  SYM theory can be seen at the three-loop order (see Ref.\cite{Kotikov:2004ss}).
The origin of such structures is related with the generalized Gribov-Lipatov reciprocity~\cite{Dokshitzer:2005bf,Dokshitzer:2006nm}.
The reciprocity function $\reciP^{\mbox{\tiny ABA}}(N)$~\cite{Dokshitzer:2005bf,Dokshitzer:2006nm,Basso:2006nk} is defined as
\begin{equation} \label{Pfunction}
\gaba(M) = \reciP^{\mbox{\tiny ABA}} \left(M+\frac{1}{2} \gaba(M) \right)\,
\end{equation}
and considerably simplifies the structure of the anomalous dimension due to additional symmetry.
Upon substituting the perturbative expansion \eqref{dimension}, one finds
\begin{equation}
\reciP^{\mbox{\tiny ABA}}(M)=\sum_{l=1}^\infty g^{2l}\,\reciP_{2l}(M)\,.\label{PABA}
\end{equation}
The reciprocity-respecting basis appears from the constraints on the harmonic polylogarithms entering into the evaluation kernel and their Mellin transformations give the necessary combinations of the harmonic sums~\cite{Dokshitzer:2006nm,Beccaria:2007bb}.
These reciprocity-respecting sums are defined recursively in the following way (see Ref.\cite{Beccaria:2009vt}):
\beqa
\Omega_a&=&S_a\,,\qquad
\Omega_{a,b}=\omega_a(\Omega_b)\,,\qquad
\Omega_{a,b,c}=\omega_a(\Omega_{b,c})\,,\ \ \ldots\,,
\\[3mm]
\omega_a(S_{b,c})&=&S_{a,b,c}-\frac{1}{2}S_{a\wedge b,c}\,,\qquad n\wedge m=\mathrm{sign}(n)\,\mathrm{sign}(m)\,(|n|+|m|)\,.
\eeqa
Several theorems were proved for the corresponding combinations of harmonic sums~\cite{Dokshitzer:2006nm,Beccaria:2007bb,Beccaria:2009vt,Beccaria:2009eq}, which allowed to reduce considerably the number of harmonic sums entering into the basis.
In this way, the result for the five-loop anomalous dimension for the twist-three operators was obtained in Ref.\cite{Beccaria:2009eq}.

The same basis can be constructed in a more elegant and natural way with the help of binomial harmonic sums.
These sums come both from the perturbative calculations of the anomalous dimensions of twist-2 operators~\cite{Moch:2004pa,Vogt:2004mw} and from the solution of the Baxter equation for the corresponding spin chain~\cite{Dippel,Kotikov:2008pv}\footnote{We will give more details about this in the forthcoming paper.}.
In our previous paper~\cite{Lukowski:2009ce} we found, that the basis from the binomial harmonic sums is equivalent to the reciprocity-respecting basis~\cite{Beccaria:2009vt}\footnote{Relations between the binomial and the nested harmonic sums together with relations between the binomial and the reciprocity-respecting harmonic sums can be found under \href{http://thd.pnpi.spb.ru/~velizh/5loop/}{\texttt{http://thd.pnpi.spb.ru/\textasciitilde velizh/5loop/}}
and  \href{http://thd.pnpi.spb.ru/~velizh/6loop3/}{\texttt{http://thd.pnpi.spb.ru/\textasciitilde velizh/6loop3/}}
} and we used the binomial harmonic sums for the calculation of five-loop anomalous dimension of the twist-2 operators both for ABA and for the wrapping correction parts.
We define the binomial harmonic sums through (see \cite{Vermaseren:1998uu})
\beq
\HBS_{i_1,\ldots,i_k}(N)=(-1)^N\sum_{j=1}^{N}(-1)^j\binom{N}{j}\binom{N+j}{j}\HS_{i_1,...,i_k}(j)\,,
\eeq
where $\HS_{i_1, \ldots ,i_k}$ are the nested harmonic sums defined in \eqref{vhs}.
One of the interesting feature of these sums is that they are defined only for \textit{positive} values of the indices $i_1,\ldots,i_k$.
More interesting, that there are binomial harmonic sums, which are expressed through the usual harmonic sums with all positive indices, for example
\beq
\HBS_{2,1,2,2,1,2,1}=8\,\HSB_{3,5,3}=8 \HS_{3,5,3}-4\,\HS_{3,8}-4\,\HS_{8,3}+2\,\HS_{11}\,.
\eeq
Namely these sums will form the basis for reciprocity functions $\reciP_i$ of ABA part for twist-3 anomalous dimension! The number of such sums, which will form the basis, for transcendentality level $k=1$, 3, 5, 7, 9, 11 are equal to 1, 2, 5, 13, 34 and 89 respectively\footnote{These numbers are the subset of the Fibonacci numbers: 0, 1, 1, 2, 3, 5, 8, 13, 21, 34, 55, 89...}, instead of $4^{k-1}$ as for the usual harmonic sums with all positive indices.
Thus, the basis is significantly reduced and the reciprocity functions $\reciP_i$ can be easily calculated.

From Eqs.(\ref{Pfunction}) and~(\ref{PABA}) one can find, that
at six-loop order the reciprocity function $\reciP_{12}$ is related with the anomalous dimension $\gamma_{12}$ (see Appendix B of Ref.\cite{Beccaria:2009eq} for five loops):
\beqa
\reciP_{12}(M)&=&\reciP_{12}\ =\
\reciP_{12}^{\rm rational}
+\reciP_{12}^{\zt}\zt
+\reciP_{12}^{\zf}\zf
+\reciP_{12}^{\zs}\zs\,,\\[2mm]
\reciP_{12}^{\rm rational}&=&
\gamma_{12}^{\rm rational}
-\frac{1}{8} \left(\gamma_6^2+2\, \gamma_4 \gamma_8^{\rm rational}
+2\, \gamma_2 \gamma_{10}^{\rm rational}\right)'\nonumber\\
&&+\frac{1}{96} \left(\gamma_4^3+6\, \gamma_2 \gamma_6 \gamma_4+3\, \gamma_2^2 \gamma_8^{\rm rational}\right)''\nonumber\\
&&-\frac{1}{768} \gamma_2^2 \left(3\, \gamma_4^2+2\, \gamma_2 \gamma_6\right)'''
+\frac{1}{6144}\left(\gamma_2^4 \gamma_4\right)''''
-\frac{1}{737280}\left(\gamma _2^6\right)'''''\,,
\label{P12}\\
\reciP_{12}^{\zt}&=&
\gamma_{12}^{\zt}
-\frac{1}{4} \left(\gamma_4 \gamma_8^{\zt}+\gamma_2 \gamma_{10}^{\zt}\right)'
+\frac{1}{32} \left(\gamma_2^2 \gamma_8^{\zt}\right)''\,,
\label{Pzt}\\
\reciP_{12}^{\zf}&=&
\gamma_{12}^{\zf}
-\frac{1}{4} \left(\gamma_2 \gamma_{10}^{\zf}\right)'\,,
\label{Pzf}\\
\reciP_{12}^{\zs}&=&
\gamma_{12}^{\zs}\,.
\label{Pzs}
\eeqa
where each prime marks derivative over $M$.

We have found the following result for $\reciP_{12}(M)$
\beqa
{\reciP^{\textrm{rational}}_{12}}&=&
-2532 \,\HBS_{2,2,2,2,2,1}
+2160 \,\HBS_{1,1,2,2,2,2,1}
+1408 \,\HBS_{1,2,1,2,2,2,1}
+1056 \,\HBS_{1,2,2,1,2,2,1}\nonumber\\ & &
+1264 \,\HBS_{1,2,2,2,1,2,1}
+1856 \,\HBS_{1,2,2,2,2,1,1}
+1248 \,\HBS_{2,1,1,2,2,2,1}
+768 \,\HBS_{2,1,2,1,2,2,1}\nonumber\\ & &
+944 \,\HBS_{2,1,2,2,1,2,1}
+1280 \,\HBS_{2,1,2,2,2,1,1}
+784 \,\HBS_{2,2,1,1,2,2,1}
+832 \,\HBS_{2,2,1,2,1,2,1}\nonumber\\ & &
+1136 \,\HBS_{2,2,1,2,2,1,1}
+864 \,\HBS_{2,2,2,1,1,2,1}
+1136 \,\HBS_{2,2,2,1,2,1,1}
+1504 \,\HBS_{2,2,2,2,1,1,1}\nonumber\\ & &
-1376 \,\HBS_{1,1,1,1,2,2,2,1}
-896 \,\HBS_{1,1,1,2,1,2,2,1}
-1120 \,\HBS_{1,1,1,2,2,1,2,1}
-1312 \,\HBS_{1,1,1,2,2,2,1,1}\nonumber\\ & &
-736 \,\HBS_{1,1,2,1,1,2,2,1}
-960 \,\HBS_{1,1,2,1,2,1,2,1}
-1088 \,\HBS_{1,1,2,1,2,2,1,1}
-640 \,\HBS_{1,1,2,2,1,1,2,1}\nonumber\\ & &
-960 \,\HBS_{1,1,2,2,1,2,1,1}
-1088 \,\HBS_{1,1,2,2,2,1,1,1}
-512 \,\HBS_{1,2,1,1,1,2,2,1}
-864 \,\HBS_{1,2,1,1,2,1,2,1}\nonumber\\ & &
-960 \,\HBS_{1,2,1,1,2,2,1,1}
-656 \,\HBS_{1,2,1,2,1,1,2,1}
-848 \,\HBS_{1,2,1,2,1,2,1,1}
-1056 \,\HBS_{1,2,1,2,2,1,1,1}\nonumber\\ & &
-320 \,\HBS_{1,2,2,1,1,1,2,1}
-768 \,\HBS_{1,2,2,1,1,2,1,1}
-992 \,\HBS_{1,2,2,1,2,1,1,1}
-768 \,\HBS_{1,2,2,2,1,1,1,1}\nonumber\\ & &
-256 \,\HBS_{2,1,1,1,1,2,2,1}
-800 \,\HBS_{2,1,1,1,2,1,2,1}
-704 \,\HBS_{2,1,1,1,2,2,1,1}
-624 \,\HBS_{2,1,1,2,1,1,2,1}\nonumber\\ & &
-720 \,\HBS_{2,1,1,2,1,2,1,1}
-928 \,\HBS_{2,1,1,2,2,1,1,1}
-512 \,\HBS_{2,1,2,1,1,1,2,1}
-672 \,\HBS_{2,1,2,1,1,2,1,1}\nonumber\\ & &
-864 \,\HBS_{2,1,2,1,2,1,1,1}
-864 \,\HBS_{2,1,2,2,1,1,1,1}
-96 \,\HBS_{2,2,1,1,1,1,2,1}
-512 \,\HBS_{2,2,1,1,1,2,1,1}\nonumber\\ & &
-864 \,\HBS_{2,2,1,1,2,1,1,1}
-864 \,\HBS_{2,2,1,2,1,1,1,1}
-640 \,\HBS_{2,2,2,1,1,1,1,1}
+1024 \,\HBS_{1,1,1,1,1,2,1,2,1}\nonumber\\ & &
+384 \,\HBS_{1,1,1,1,1,2,2,1,1}
+640 \,\HBS_{1,1,1,1,2,1,1,2,1}
+768 \,\HBS_{1,1,1,1,2,1,2,1,1}
+512 \,\HBS_{1,1,1,1,2,2,1,1,1}\nonumber\\ & &
+384 \,\HBS_{1,1,1,2,1,1,1,2,1}
+640 \,\HBS_{1,1,1,2,1,1,2,1,1}
+832 \,\HBS_{1,1,1,2,1,2,1,1,1}
+256 \,\HBS_{1,1,1,2,2,1,1,1,1}\nonumber\\ & &
+192 \,\HBS_{1,1,2,1,1,1,1,2,1}
+512 \,\HBS_{1,1,2,1,1,1,2,1,1}
+832 \,\HBS_{1,1,2,1,1,2,1,1,1}
+832 \,\HBS_{1,1,2,1,2,1,1,1,1}\nonumber\\ & &
+256 \,\HBS_{1,2,1,1,1,1,2,1,1}
+704 \,\HBS_{1,2,1,1,1,2,1,1,1}
+832 \,\HBS_{1,2,1,1,2,1,1,1,1}
+640 \,\HBS_{1,2,1,2,1,1,1,1,1}\nonumber\\ & &
+448 \,\HBS_{2,1,1,1,1,2,1,1,1}
+704 \,\HBS_{2,1,1,1,2,1,1,1,1}
+640 \,\HBS_{2,1,1,2,1,1,1,1,1}
+640 \,\HBS_{2,1,2,1,1,1,1,1,1}\nonumber\\ & &
-512 \,\HBS_{1,1,1,1,1,2,1,1,1,1}
\,.\label{ABArat}
\eeqa
The functions $\reciP^{\zeta(3)}_{12}$, $\reciP^{\zeta(5)}_{12}$ and $\reciP^{\zeta(7)}_{12}$ are easy to determine
\beqa
\reciP_{12}^{\zeta(3)}&\!\!=\!&
-32 \big(
45  \,\HBS_{1,2,2,2,1}\!
+21 \,\HBS_{2,1,2,2,1}\!
+19 \,\HBS_{2,2,1,2,1}\!
+39 \,\HBS_{2,2,2,1,1}\!
-6  \,\HBS_{2,1,1,1,2,1}\!
-8  \,\HBS_{2,1,1,2,1,1}\nonumber\\&&
-32 \,\HBS_{1,1,2,2,1,1}
-10 \,\HBS_{1,2,1,1,2,1}
-12 \,\HBS_{1,2,1,2,1,1}
-28 \,\HBS_{1,2,2,1,1,1}
-24 \,\HBS_{1,1,1,2,2,1}\nonumber\\&&
-20 \,\HBS_{1,1,2,1,2,1}\!
-16 \,\HBS_{2,1,2,1,1,1}\!
-12 \,\HBS_{2,2,1,1,1,1}\!
+24 \,\HBS_{1,1,1,2,1,1,1}\!
+24 \,\HBS_{1,1,2,1,1,1,1}
\big)
,\label{ABAz3}\\
\reciP_{12}^{\zeta(5)}&\!\!=\!&
-64 \left(
29 \,\HBS_{1,2,2,1}\!
+5 \,\HBS_{2,1,2,1}\!
+28 \,\HBS_{2,2,1,1}\!
-20 \,\HBS_{1,1,2,1,1}\!
-19 \,\HBS_{1,2,1,1,1}\!
+3 \,\HBS_{2,1,1,1,1}\!
\right),
\label{ABAz5}\\
\reciP_{12}^{\zeta(7)}&\!\!=\!&
-3360 \left(\HBS_{1,2,1}+\HBS_{2,1,1}\right)
\,.\label{ABAz7}
\eeqa
The expression for the six-loop anomalous dimension of twist-three operators from ABA in the canonical basis of the nested harmonic sums (\ref{vhs}) can be found on our web-page.

\section{Wrapping correction} \label{sec:wrapping}

In this section we will write explicitly all equations, which should be modified in Ref.\cite{Lukowski:2009ce} for the calculations in the case of twist-three operators. The wrapping correction is calculated by evaluating the first L\"uscher correction at weak-coupling along the lines advocated in \cite{Bajnok:2008bm,Bajnok:2008qj,Bajnok:2009vm}.
The interested reader should read this subsection together with our previous paper \cite{Lukowski:2009ce}, where all definitions and a lot of details can be found.

First of all, the length of twist-three operator is equal to $L=3$, which according to general formulae for F-term (see Subsection {\bf 5.2} in Ref.~\cite{Bajnok:2009vm} for details)
\beq
\Delta^F_{w}=-\sum_{Q=1}^{\infty}\int_{-\infty}^{\infty}\!\frac{dq}{2\pi}
\left(\frac{
z^{-}}{z^{+}}\right)^{L}\sum_{b}(-1)^{F_{b}}\left[S_{Q-1}(q,u_{i})S_
{Q-1}(q,u_{ii})\right]_{b(11)}^{b(11)}\label{wrapping}
\eeq
renders the exponential part to be of the form
\beq
\left(
\frac{z^-}{z^+}\right)^3=\frac{64g^4}{(q^{2}+Q^{2})^{3}}\left[1-g^{2}\frac{24}{(q^{2}+Q^{2})}\right]\,.
\label{modification}
\eeq
This leads to the following modification of the last factor in Eq.(5.6) of~\cite{Lukowski:2009ce}
\begin{equation}
\frac{16}{(q^{2}+Q^{2})^{2}}\to\frac{64}{(q^{2}+Q^{2})^{3}}
\end{equation}
and
\begin{equation}
\label{e.integral}
\intd^{(10,0)}(M,q)=64\, S_{1}^{2}\,
\frac{T_M(q,Q)^{2}}{R_M(q,Q)}\frac{64}{(q^{2}+Q^{2})^{3}}\,,
\end{equation}
where $\intd^{(10,0)}(M,q)$ is the integrand for the five-loop wrapping corrections to the anomalous dimension of twist-3 operators~\cite{Beccaria:2009eq}
\begin{equation}
\label{5integral}
\Delta_w^{(10)}(M)=-\sum_{Q=1}^{\infty}\int_{-\infty}^{\infty}\!\frac{dq}{2\pi}
\ \intd^{(10,0)}(M,q)\,,
\end{equation}
with $T_M(q,Q)$ and $R_M(q,Q)$ are given by
\begin{eqnarray}
R^{(0)}_M(q,Q)&=&
\Baxter\left(\frac{q - i(-1 + Q)}{2}\right)
\Baxter\left(\frac{q + i(-1 + Q)}{2}\right)
\Baxter\left(\frac{q - i(1 + Q)}{2}\right)\times\nonumber\\
&&\times\Baxter\left(\frac{q + i(1 + Q)}{2}\right)\,,\\
T^{(0)}_M(q,Q)&=&\!\sum_{j=0}^{Q-1}\!
\left( \frac{1}{2 j - i q - Q} -\frac{(-1)^M}{2 (j + 1) - i q - Q} \right)
\!\Baxter\!\!\left(\frac{q - i (Q - 1)}{2}  + i j\right)\!.
\end{eqnarray}
Farther, the one-loop energy of twist-three operators differs from the twist-two one. It is given by
\beq
    \sum_{k=1}^{M}\frac{16}{1+4u_{k}^{2}}=8S_{1}\big(\tfrac{M}{2}\big)
\eeq
and we should replace everywhere in $S_1(M)$ argument $M$ to $M/2$. Moreover,
the one-loop Baxter function Eq.(5.9) of~\cite{Lukowski:2009ce}, for twist-three case is given by \cite{Kotikov:2007cy,Beccaria:2007cn}
\begin{equation}
\label{LOBaxter}
\Baxter(u)={}_4 F_3\left(\left. \begin{array}{c}
    -\frac{M}{2}, \ \frac{M}{2}+1,\ \frac{1}{2}+iu ,\ \frac{1}{2}-iu\\[3mm]
    1,\  1 ,\ 1 \end{array}
    \right| 1\right) \,.
\end{equation}

The integrand $Y^{(12,0)}$ for the six-loop wrapping corrections to the anomalous dimension of twist-3 operators can be split by analogy with the five loops for the twist-2 operators into the matrix part, scalar part, the exponential term and the dressing factor (see details in Refs.\cite{Bajnok:2009vm} and \cite{Lukowski:2009ce}).

For the scalar part of $Y^{(12,0)}$ it is easy to find, that $S_{-2}(M)$ term in Eq.(5.11) of~\cite{Lukowski:2009ce} drop out (and do not forget to change the argument of $S_1$ in the last term)
\beq
\frac{S_{0rat}^{(2)}(M,q,Q)}{\sca^{(0)}(M,q,Q)}=\frac{4\,q}{q^2+Q^2}\left( \frac{\DBaxter(\frac{-q - i(1 + Q)}{2})}{\Baxter (\frac{-q - i(1 + Q)}{2})}
- \frac{ \DBaxter (\frac{q - i(1 + Q)}{2})}{ \Baxter (\frac{(q - i(1 + Q)}{2})}\right)
- \frac{16\,Q\,S_{1}\!\big(\tfrac{M}{2}\big)}{q^2 + Q^2}\,.
\eeq

The exponential part Eq.(5.15) of~\cite{Lukowski:2009ce} should be modified according to Eq.(\ref{modification}) as
\begin{equation}
\frac{\ex^{(6)}(q,Q)}{\ex^{(4)}(q,Q)}=-\frac{24}{q^2 + Q^2} \,.
\end{equation}

The two-loop solution $\Baxtersub(u)$ Eq.(5.25) of~\cite{Lukowski:2009ce} can be taken for twist-3 from Ref.\cite{Kotikov:2008pv}
\begin{eqnarray}\label{Q1full}
\Baxtersub(u)&=&\left(2S_1(\tfrac{M}{2})\left( 4S_1(\tfrac{M}{2})-2S_1(M)\right)+3\,S_2(\tfrac{M}{2})\right)
\Baxter(u)
\nonumber\\
&&+ \, 2S_1(\tfrac{M}{2})\, \frac{\p}{\p\delta} \, {}_4 F_3 \left. \left(\left.
    \begin{array}{c}
        -\tfrac{M}{2}, \ \tfrac{M}{2}+1+2\delta,\ \tfrac{1}{2}+iu ,\ \tfrac{1}{2}-iu\\
        1+\delta,\  1+\delta ,\ 1 \end{array}
    \right| 1\right)\right|_{\delta=0} \nonumber\\
&&- \, \phantom{\gamma}\frac{3}{2} \, \frac{\p^2}{\p\delta^2} \, {}_4 F_3 \left. \left(\left.
    \begin{array}{c}
        -\tfrac{M}{2}, \ \tfrac{M}{2}+1,\ \tfrac{1}{2}+iu ,\ \tfrac{1}{2}-iu\\
        1+\delta,\  1-\delta ,\ 1 \end{array}
    \right| 1\right)\right|_{\delta=0}\,.
\end{eqnarray}

In the modification of ABA part it is necessary to change the powers in Eq.(5.32) of~\cite{Lukowski:2009ce}
\beq
BY(u_k)=\left(\frac{u_k+\frac{i}{2}}{u_k-\frac{i}{2}}\right)^3\prod^M_{j \neq k} \frac{u_k-u_j+i}{u_k-u_j-i}
\eeq
and in Eq.(5.33) of~\cite{Lukowski:2009ce}
\beq \label{phi}
\Phi_k=\sum^\infty_{Q=1}\int^{\infty}_{-\infty} \frac{dq}{2\pi i} \left(\frac{z^-}{z^+}\right)^3\,\textrm{str} \bigg\{ S_Q(q,u_1)\ldots \partial S_Q(q,u_k)\ldots S_Q(q,u_M) \bigg\}\,.
\eeq

We have checked, that the six-loop integrand obtained with formulae from Section 5 of Ref.\cite{Lukowski:2009ce}, modified according to this section, gives zero for the sum over dynamical poles in $M=2$ and $M=4$ cases\footnote{I thank Tomasz \L ukowski and Adam Rej for the check of the cancelation.}.

\subsection{Calculation and result}
For calculations we use the same method as in Ref.\cite{Lukowski:2009ce}.
With formulae from Section 5 of Ref.\cite{Lukowski:2009ce}, modified according to the previous section, we obtain the integrand for given $M$, put $Q=1$, perform numerical integration over $q$ and repeat this procedure for $Q=2, 3 , \dots$.
Then, the high-precision result is used to find the coefficients in the following transcendental basis
\beq
\left\{\zeta(9), \zeta(3)^3, \zeta(5)\, \zeta(3), \zeta(7), \zeta(3)^2, \zeta(5), \zeta(3), 1\right\}
\eeq
using {\texttt{EZ-Face}}~\cite{EZFace} and/or {\texttt{PSLQ}} algorithm~\cite{PSLQ} implemented on {\texttt{MATHEMATICA}}~\cite{PSLQMATH} ({\texttt{EZ-Face}} is limited with 100 digits of accuracy). This allowed us to conjecture the form of the functions multiplying zeta functions. For higher values of $M$ the non-rational part can be subtracted from the full result and it is sufficient to perform computations to the accuracy that allows to rationalize the result. In this way we obtain 25 values up to $M=50$.

For the reconstruction of the full $M$-dependence we assume that the wrapping corrections preserve the reciprocity symmetry. This implies that a part of the wrapping correction may be found from the lower order results
\beq \label{gamma10P}
\Delta^{(12)}_{w}=\frac{1}{2} \left(\Delta_w^{(10)}\gamma_2\right)'
+\reciP_{12}^{w}\,,
\eeq
where $\gamma_2$ is the one-loop anomalous dimension and $ \Delta_w^{(10)}$ is the five-loop wrapping correction found in~\cite{Beccaria:2009eq} (all harmonic sums being of argument $M/2$)
\beqa
\gamma_2(M)&=&\reciP_2(M)=8\,\HS_1(M/2)=4\,\HBS_1(M/2)\,,\\
 \Delta_w^{(10)}&=&\reciP_{10}^{w}=
-64 \,g^{10}\,\HS_{1}^{2}\Big( 35\zeta(7) -40\HS_{2}\zeta(5) +(-8\HS_{4}+16\HS_{2,2})\zeta(3) \nonumber\\
&&+2\HS_{7}-4\HS_{2,5}-2\HS_{3,4}-4\HS_{4,3}-2\HS_{6,1}+8\HS_{2,2,3}+4\HS_{3,3,1}\Big)\label{P10}\\
&&=-64\,g^{10}\,\Omega_1^2\,\Big(35\,\zeta(7)+4\,\Omega_{3,3,1}+8\,\underline{\Omega_{2,2,3}}+24\,\zeta(3)\,\underline{\Omega_{2,2}}-\Omega_7\Big)\,.
\label{P10om}
\eeqa
Here $\underline{\Omega_{\mathbf m}}$ are the complementary reciprocity-respecting harmonic sums defined through complementary harmonic sums $\underline{\HS_{\mathbf m}}$ as (see Appendix A in Ref.\cite{Beccaria:2007bb} and Ref.\cite{Beccaria:2009vt})
\beqa
\underline{\HS_{a, {\mathbf {m}}}}(M)=
\HS_{a, {\mathbf{m}}}(M)-
\HS_{a}(M)\underline{\HS_{\mathbf{m}}}(\infty)\,.
\eeqa

We suggest, that the remaining part $\reciP_{12}^w$ should have the same general structure as for twist-two case~\cite{Lukowski:2009ce}
\beqa \label{gamma10split}
\reciP_{12}^{w}&=&2\,\reciP_2^2 \reciT
+2\,\reciP_2 \Big(2\,\reciP_4 +\frac{1}{16} \reciP^3_2\Big)
\Big( 35\zeta(7) -40\HS_{2}\zeta(5) +(-8\HS_{4}+16\HS_{2,2})\zeta(3) \nonumber\\
&&+2\HS_{7}-4\HS_{2,5}-2\HS_{3,4}-4\HS_{4,3}-2\HS_{6,1}+8\HS_{2,2,3}+4\HS_{3,3,1}\Big)\,, \label{Ttilde}
\eeqa
with $\reciT$ and $\reciP_{4}$ are given by
\beqa
\reciT&=&
\zn\, \reciT_{\zn}
+{\zt}^3\, \reciT_{{\zt}^3}
+\zf\,\zt\, \reciT_{\zf\zt}
+\zs\, \reciT_{\zs}
+{\zt}^2\, \reciT_{{\zt}^2}\nonumber\\
&&+\zf\, \reciT_{\zf}
+\zt\, \reciT_{\zt}
+\reciT_{\rm rational}\,,\label{reciT}\\
\reciP_{4}&=&-8\HS_3\,.\label{P4}
\eeqa

Applying the principle of maximal transcendentality~\cite{Kotikov:2002ab} we conclude
that the transcendentality of the components
$\reciT_{\zn}$, $\reciT_{\zt^3}$, $\reciT_{\zf\zt}$, $\reciT_{\zs}$, $\reciT_{\zt^2}$, $\reciT_{\zf}$, $\reciT_{\zt}$, $\reciT_{\rm rational}$ should be equal to 2, 2, 3, 4, 5, 6, 8 and 11 respectively. The lowest-transcendentality functions,
$\reciT_{\zn}$, $\reciT_{\zt^3}$, $\reciT_{\zf\zt}$, $\reciT_{\zs}$ and $\reciT_{\zt^2}$ may be obtained from first few values 
\beqa
\reciT_{\zn}&=& 64512\ \HS_1^2 \,,\label{Tzn}\\
\reciT_{\zt^3}&=& 0\,,\label{Tzt3}\\
\reciT_{\zf\zt}&=& -7680\ \HS_1^3\,,\label{Tzfzt}\\
\reciT_{\zs}&=& -62720\ \HS_1^2 \HS_2\,,\label{Tzs}\\
\reciT_{\zt^2}&=& 6144\ \HS_1^3 \HS_2\,.\label{Tzt2}
\eeqa

Unfortunately, as can be seen from Eq.(\ref{P10}) the basis for other $\reciT_i$ will contain the harmonic sums with even numbers among indices, in contrast to the ABA part.
Thus, the basis will extended considerably, even if one take into account the reciprocity-respecting parity even sums~\cite{Beccaria:2009eq}.
Moreover, the sums are appeared, which depend on $M$, not on $M/2$.
This can be seen by applying \texttt{MATHEMATICA} function \texttt{FactorInteger} to the denominators of the obtained values. For example, for $M=8$ we have found:
\beqa
\Delta_w^{(12)}(8)&=&
280000\, \zeta (9)
-\frac{625000}{9}\, \zeta (3) \zeta (5)
-\frac{1386875}{3}\,  \zeta (7)
+\frac{6406250}{81}\,  \zeta (3)^2\nonumber\\
&&+\frac{9539440625}{40824}\, \zeta (5)
-\frac{294429940625}{2939328}\, \zeta (3)
+\frac{277950256725625}{6772211712}
\eeqa
and \texttt{FactorInteger} gives for the denominator of $\zeta(5)$'s coefficient:
\beq
40824=2^3\times 3^6\times 7^1\,.
\eeq
For argument of harmonic sums equal to $M/2$ the last number $7^1$ should be absent if $M=8$. Thus, the basis expands extremely. Fortunately, there is a nice observation in Appendix B of Ref.\cite{Bajnok:2008qj} about relation of the exact result with the functions $P_M$ and $T_M(q,Q)$ from the integrand for twist-two case at four loops.
We assume that similar properties will remain for the operators of twist-3.
Following Ref.\cite{Bajnok:2008qj}, we rewrite $T_M(q,Q)$ as
\beqa
T_M(q,Q) & = &
\sum_{j=0}^{Q-1}\left[\frac{1}{2j-iq-Q}-\frac{1}{2(j+1)-iq-Q}\right]
P_M\left(\frac{1}{2}(q-i(Q-1))+ij\right)\nonumber\\
&=&\frac{iP_M\left(\frac{1}{2}(q-i(Q-1))\right)}{q-iQ}
-\frac{iP_M\left(\frac{1}{2}(q+i(Q-1))\right)}{q+iQ}+\widetilde{T}_M(q,Q)\,,\\[3mm]
\widetilde{T}_M(q,Q)&=&\sum_{j=1}^{Q-1}\frac{P_M\left(\frac{1}{2}(q-i(Q-1))+
  ij\right)
-P_M\left(\frac{1}{2}(q-i(Q+1))+ij\right)}{2j-iq-Q}\,,
\eeqa
expand $\widetilde T_M(q,Q)$ over $Q$ for a lot of $M$ and
reconstruct the general dependence on M as a combination of harmonic sums for each power of $Q$.
As can be seen from Appendix~\ref{SpecialSums}, for example, in the case of $Q^1$ the harmonic sums $\HS_{-2,1}(M)$ and $\HS_{2,1}(M)$ enter into answer.
Thus, to reconstruct the full $M$-dependence for $\zeta(5)$'s term we write down the following basis (if not written explicitly, the argument of harmonic sums is $M/2$):
\beq
\HSB_1^2\{\HSB_{4},\HSB_{3,1},\HSB_{2,2},\HSB_{1,3},\HSB_{2,1,1},\HSB_{1,2,1},\HSB_1\HSB_{-2,1}(M),\HSB_1\HSB_{2,1}(M)\}
\eeq
and we have found
\beqa
\reciT_{\zeta(5)}&=&
-512 \,\HSB_1^2
\Big(
5 \,\HSB_1
\big(
8 \,\HSB_{2,1}
+\,\HSB_3
\big)
-50 \,\HSB_{2,2}
-6 \,\HSB_{3,1}
+20 \,\HSB_{2,1,1}
+15 \,\HSB_4
\Big)\nonumber\\&&
+40960 \,\HSB_1^3
\Big(
\HSB_{-2,1}(M)
+\,\HSB_{2,1}(M)
\Big)\,.\label{Tzf}
\eeqa
For $\reciT_{\zeta(3)}$ part we add to the standard basis from the harmonic sums with argument $M/2$
\beqa
&\HSB_1^2&\{\HSB_{6},\HSB_{5,1},\HSB_{4,2},\HSB_{3,3},\HSB_{2,4},\HSB_{1,5},
\HSB_{4,1,1},\HSB_{1,4,1},
\HSB_{3,2,1},\HSB_{3,1,2},
\HSB_{2,3,1},\HSB_{2,1,3},
\HSB_{1,3,2},\HSB_{1,2,3},\nonumber\\
&&\HSB_{2,2,2},
\HSB_{3,1,1,1},\HSB_{1,3,1,1},
\HSB_{2,2,1,1},\HSB_{2,1,2,1},
\HSB_{2,1,1,2},\HSB_{1,2,2,1},
\HSB_{1,2,1,2}\}
\eeqa
the following combinations of harmonic sums with argument $M$
\beqa
\MHSB_{2,2,\underline{\bf 1}}(M)&=&
\HSB_{2,2,1}(M)+\HSB_{-2,2,1}(M)+\HSB_{2,-2,1}(M)+\HSB_{-2,-2,1}(M)\,,\\
\MHSB_{2,\underline{\bf 1},2}(M)&=&
\HSB_{2,1,2}(M)+\HSB_{-2,1,2}(M)+\HSB_{2,1,-2}(M)+\HSB_{-2,1,-2}(M)
\eeqa
multiplied by $\HSB_1^3(M/2)$. With this basis we have obtained
\beqa
\reciT_{\zeta(3)}&=&
1024 \,\HSB_1^2
\Big(
\,\HSB_1
\big(
8\big(
\,\HSB_{2,1,2}
+3 \,\HSB_{2,2,1}
\big)
-\HSB_5
\big)
-\,\HSB_{5,1}
+2 \,\HSB_{2,4}
+3 \,\HSB_{4,2}
+4 \,\HSB_{2,1,1,2}\nonumber\\&&
+4 \,\HSB_{2,1,2,1}
+8 \,\HSB_{2,2,1,1}\Big)
-32768\, \HSB_1^3
\big(\MHSB_{2,\underline{\bf 1},2}(M)+3\,\MHSB_{2,2,\underline{\bf 1}}(M)\big)\,.\label{reciTz3}
\eeqa

To reconstruct the rational part we need the basis with at least 83 reciprocity-respecting harmonic sums with argument $M/2$ and at least two following combinations of the rec\-i\-proc\-i\-ty-respecting harmonic sums with argument $M$ (see details in Appendix~\ref{SpecialSums}):
\beqa
\MHSB_{2,2,\underline{\bf 1},3}(M)&=&
\HSB_{2,2,1,3}(M)+\HSB_{-2,2,1,3}(M)+\HSB_{2,-2,1,3}(M)+\HSB_{2,2,1,-3}(M)+\HSB_{-2,-2,1,3}(M)\nonumber\\&&
+\HSB_{2,-2,1,-3}(M)+\HSB_{-2,2,1,-3}(M)+\HSB_{-2,-2,1,-3}(M)\,,\\
\MHSB_{2,\underline{\bf 1},2,3}(M)&=&
\HSB_{2,1,2,3}(M)+\HSB_{-2,1,2,3}(M)+\HSB_{2,1,-2,3}(M)+\HSB_{2,1,2,-3}(M)+\HSB_{-2,1,-2,3}(M)\nonumber\\&&
+\HSB_{2,1,-2,-3}(M)+\HSB_{-2,1,2,-3}(M)+\HSB_{-2,1,-2,-3}(M)\,.
\eeqa
So, we should calculate $85$ values up to $M=170$.
This is far beyond computational capabilities.
However, as we well know from the previous results, coefficients in the front of the reciprocity-respecting harmonic sums should be integers with a lot of zeros among them.
We can try to find these coefficients with the Number Theory.
The most simple algorithm for the solution of given problem  is the LLL-algorithm~\cite{Lenstra:1982}, which is realized in \texttt{MATHEMATICA} with function
\texttt{LatticeReduce}.
We calculate the values of all 85 reciprocity-respecting and special sums in the basis~\ref{RRSums9}-\ref{SpecSums9B} up to $M=48$.
So, we have $24$ equations in the linear system for $85$ variables together with our corresponding results with subtracted low order anomalous dimension according to Eq.(\ref{gamma10P}).
We eliminate $23$ variables and we remain with one equation on $62$ variables.
According to the realization of LLL-algorithm\footnote{see {\bf Application} on
\href{http://reference.wolfram.com/mathematica/ref/LatticeReduce.html}{\texttt{{http://reference.wolfram.com/mathematica/ref/LatticeReduce.html}}}} we replace the last column in $64\times 64$ unity matrix with the last equation of our system ($62 + 1$ numbers) and zero as the last element.
After four minutes\footnote{The computation of $M=50$ requires about 200 hours.} of calculations \texttt{MATHEMATICA} gives:
\beqa
\reciT_{\rm{rational}}&=&
-512 \,\HSB_1^2
\Big(
\HSB_1
\big(
\HSB_{2,6}
+\HSB_{5,3}
-8 \big(
\HSB_{2,1,2,3}
+\HSB_{2,2,1,3}
\big)
\big)
+\,\HSB_9
+\HSB_{2,1,6}
+\HSB_{2,6,1}
-\HSB_{3,3,3}\nonumber\\&&
-\HSB_{3,5,1}
+\HSB_{5,1,3}
-\HSB_{5,3,1}
+\HSB_{7,1,1}
-2 \,\HSB_{2,2,5}
-2 \,\HSB_{2,4,3}
-3\,\HSB_{4,2,3}
-4\,\HSB_{2,1,1,2,3}\nonumber\\&&
-4 \,\HSB_{2,1,2,1,3}
-4 \,\HSB_{2,1,2,3,1}
-4 \,\HSB_{2,2,1,3,1}
-4 \,\HSB_{3,1,1,3,1}
-4 \,\HSB_{3,3,1,1,1}
-8 \,\HSB_{2,2,1,1,3}\nonumber\\&&
-8 \,\HSB_{2,2,3,1,1}
\Big)
-65536\, \HSB_1^3
\Big(
\MHSB_{2,\underline{\mathbf{1}},2,3}(M)
+\MHSB_{2,2,\underline{\mathbf{1}},3}(M)
\Big)\,.\label{reciTR}
\eeqa
Note, that \texttt{LatticeReduce} gives $64$ possible solutions but all other solutions contain large integers and only few zeros.
We have checked the obtained result for $M=25$ and have found the full agreement, which can serve as confirmation of the correctness for obtained result.

Now we are ready go to the complementary functions $\underline{\HSB_{\bf{a}}}$ for testing of $\reciT$ on the parity following Ref.\cite{Beccaria:2009eq}.
We have found
\beqa
\underline{\reciT_{\zn}}&=& 64512\ \HSB_1^2 \,,\\
\underline{\reciT_{\zt^3}}&=& 0\,,\\
\underline{\reciT_{\zf\zt}}&=& -7680\ \HSB_1^3\,,\\
\underline{\reciT_{\zs}}&=& 0\,,\\
\underline{\reciT_{\zt^2}}&=& 0\,,\\
\underline{\reciT_{\zeta(5)}}&=&
-1536 \,\HSB_1^2
\Big(
5 \,\HSB_1
\HSB_3
-19 \,\underline{\HSB_{2,2}}
-\HSB_{3,1}
\Big)\,,
\eeqa
where we don't write the expressions for $\underline{\reciT_{\zeta(3)}}$ and $\underline{\reciT_{\rm{rational}}}$ as the harmonic sums,
entering into Eqs.(\ref{reciTz3}) and (\ref{reciTR}) have only odd numbers through indices or two even numbers with other odd numbers through indices.
According to theorem from Ref.\cite{Beccaria:2009eq} all complementary functions in the expressions for $\underline{\reciT_i}$ are parity-even,
then $\underline{\reciT}$ is parity-even also and satisfy reciprocity.
With help of the \texttt{SUMMER}~\cite{Vermaseren:1998uu} and \texttt{HARMPOL}~\cite{Remiddi:1999ew}
packages for \texttt{FORM}~\cite{Vermaseren:2000nd} and the \texttt{HPL} package~\cite{Maitre:2005uu}
for \texttt{MATHEMATICA}, we checked, that the combinations of harmonic sums in the special sums Eqs.(\ref{reciTz3}) and (\ref{reciTR})
translated into polylogarithms satisfy Gribov-Lipatov relation~\cite{Dokshitzer:2006nm} up to $\zeta$'s terms.
However, the complete analysis of the full result for the six-loop anomalous dimension of the twist-3 operators is unavailable
with these programs due to their limitations. Nevertheless it seems, that the reciprocity will work for the full result also.

\section{Large $M$ asymptotic and analytical continuation}
In this section we check our result for the six-loop anomalous dimension of the twist-three operators against the known constraints.
The $M\to\infty$ limit can be easily calculated with use of the \texttt{SUMMER} package~\cite{Vermaseren:1998uu} for \texttt{FORM} ~\cite{Vermaseren:2000nd}
\beq
\lim_{M \to \infty} \Delta_w^{(12)} (M)=0\,.
\eeq
This means that, the wrapping effects do not influence the scaling function.

Moreover, there are predictions from the empirically established resummation formulae for the analytical continuation to $M=-2+\omega$ from Ref.\cite{Kotikov:2007cy}
\beq
\gamma=-8\,\frac{g^2}{\omega}\left(\frac{1}{1-t}-\zeta(2)\,\frac{1+3 t^2}{(1-t)^2}\,\omega^2\right),\qquad t=\frac{g^2}{\omega^2}\,,
\eeq
which give at six loops:
\beq
\gamma_{12}=-\frac{8}{\omega^{11}}+\frac{144}{\omega^9}\zeta(2)\,.\label{Ressum}
\eeq

With help of the \texttt{SUMMER}~\cite{Vermaseren:1998uu} and \texttt{HARMPOL}~\cite{Remiddi:1999ew}
packages for \texttt{FORM}~\cite{Vermaseren:2000nd} and the \texttt{HPL} package~\cite{Maitre:2005uu}
for \texttt{MATHEMATICA}, we have found the following results for the analytical continuation to $M=-2+\omega$ from ABA part,
from the wrapping corrections part without special sums and from the special sums (the last terms) in the wrapping corrections correspondingly:
\beqa
\gamma_{12}^{\mathrm{ABA}}=-\frac{1288}{\omega^{11}}+\frac{6800}{\omega^9}\,\zeta(2)\,,\qquad
\Delta_{12}^{w}=\frac{768}{\omega^{11}}-\frac{5120}{\omega^9}\,\zeta(2)
+\frac{512}{\omega^{11}}-\frac{1536}{\omega^9}\,\zeta(2)\,.
\eeqa
Summing up all three pieces we obtain full agreement with (\ref{Ressum}), which can serve as additional test for the correctness of obtained result.

In the end we want to write explicitly few first values of the six-loop anomalous dimension of the twist-3 operators from our general result~(\ref{P12})-(\ref{ABAz7}), (\ref{gamma10P})-(\ref{Tzt2}), (\ref{Tzf}), (\ref{reciTz3}) and (\ref{reciTR}). The anomalous dimension of the most simple operator with $M=2$, which is the analog of Konishi operator in twist-2 case, is equal to
\beqa
\gamma_{12}(M=2)&=&
-53016
-18176 \,\zt
-16128 \,\zf
-13440 \,\zs\nonumber\\&&
+3840
+6144 \,\zt^2
-7680 \,\zf \,\zt
+1024 \,\zt
-3584 \,\zf
-53760 \,\zs\nonumber\\&&
+64512 \,\zn\,,
\eeqa
where the first line is the contribution from ABA. In the sum it gives
\beqa
\gamma_{12}(M=2)&=&
-49176
+6144 \,\zt^2
-7680 \,\zf \,\zt
-17152 \,\zt
-19712 \,\zf\nonumber\\&&
-67200 \,\zs
+64512 \,\zn\,.
\eeqa
For $M=4$ we have found:
\beqa
\gamma_{12}(M=4)&=&
-\frac{31468863}{256}
-36018 \,\zt
-29916 \,\zf
-22680 \,\zs\nonumber\\&&
+\frac{127431}{8}
+25920 \,\zt^2
-25920 \,\zf \,\zt
-14364 \,\zt
+36216 \,\zf\nonumber\\&&
-176400 \,\zs
+145152 \,\zn\,.
\eeqa
And for $M=6$, which ABA part (divided by $2^{12}$) can be found in Ref.\cite{Belitsky:2009mu}, we have:
\beqa
\gamma_{12}(M=6)&=&
-\frac{2720281112987}{15116544}
-\frac{317437583}{6561} \,\zt
-\frac{3175148}{81} \,\zf
-\frac{773080}{27} \,\zs\nonumber\\&&
+\frac{68920603499}{2361960}
+\frac{4174016}{81} \,\zt^2
-\frac{425920}{9}\,\zf \,\zt
-\frac{1650710908}{32805} \,\zt\nonumber\\&&
+\frac{87869320}{729} \,\zf
-\frac{8562400}{27} \,\zs
+216832\,\zn\,.
\eeqa


 \subsection*{Acknowledgments}
I would like to thank to Tomasz \L ukowski and Adam Rej for the nice collaboration during working on the calculations of the five-loop anomalous dimension for the twist-two operators. I would like to thank them on the earlier stage of presented calculations. This work is supported by RFBR grants 10-02-01338-a, RSGSS-65751.2010.2. 

\appendix
\section{Special sums} \label{SpecialSums}

In this Appendix we will show the appearance of the special harmonic sums in the basis for the reconstruction of the general expression
for the wrapping corrections.
In Appendix B of Ref.\cite{Bajnok:2008qj} it is shown, that the results for some of $\zeta$'s contributions to the four-loop anomalous dimension
of the twist-two operators from the wrapping corrections can be obtained exactly with the expansion over $Q$
of the functions $P_M$ and $T_M(q,Q)$ entering into integrand.
In our case we will concentrate only on $T_M(q,Q)$ function, which can be written according to~\cite{Bajnok:2008qj} in the following form
\beqa
T_M(q,Q) & = &
\sum_{j=0}^{Q-1}\left[\frac{1}{2j-iq-Q}-\frac{1}{2(j+1)-iq-Q}\right]
P_M\left(\frac{1}{2}(q-i(Q-1))+ij\right)\nonumber\\
&=&\frac{iP_M\left(\frac{1}{2}(q-i(Q-1))\right)}{q-iQ}
-\frac{iP_M\left(\frac{1}{2}(q+i(Q-1))\right)}{q+iQ}+\widetilde{T}_M(q,Q)\,,\\[3mm]
\widetilde{T}_M(q,Q)&=&\sum_{j=1}^{Q-1}\frac{P_M\left(\frac{1}{2}(q-i(Q-1))+
  ij\right)
-P_M\left(\frac{1}{2}(q-i(Q+1))+ij\right)}{2j-iq-Q}\,.
\eeqa
We substitute two-loop Baxter function $P_M^{(2)}$ from Eq.(\ref{Q1full}) instead of $P_M$,
expand $\widetilde T_M(q,Q)$ over $Q$ for a lot of $M$ and for each power of $Q$ reconstruct the combination of harmonic sums,
which give expression for arbitrary $M$.
For $Q=0$ the transcendentality level for obtained result will be the same, as transcendentality level of two-loop Baxter function,
that is equal to $3$ and we can easily found (argument of harmonic sums is $M/2$ if not written explicitly)
\beq
Q^0\ :\quad\ 6 \HS_3-2 \HS_1 \left(-4 S_1(M) \HS_1+8 \HS_1^2+\HS_2\right)\,.
\eeq
Similar structures come from the expansion of $P_M$, so the harmonic sums with argument $M$ will drop out in the expression for the anomalous dimension.
For the first power of $Q$ the transcendentality level for obtained result is $4$ and we have found
\beqa
Q^1&:\quad&2 \left(-12 \HS_{1,3}+3 \HS_{2,2}-8 \HS_{3,1}+7 \HS_{1,1,2}+3
   \HS_{1,2,1}-\HS_{2,1,1}+4 \HS_1^4+4 \HS_4\right)\nonumber\\&&
-4 S_1(M) \HS_1 \left(\HS_1^2+\HS_2\right)+16 \HS_1 \left(S_{-2,1}(M)+S_{2,1}(M)\right)\,.
\eeqa
The sums in the last term are the special one, which do not appear in the expansion of $P_M$ and can enter into the anomalous dimension.
So, we will add such sums multiplied by common factor $S_1^2(M/2)$ to the basis for the transcendently level 6, which is relevant for the $\zeta(5)$ contribution.
For the second power of $Q$ the result of expansion $\widetilde T_M(q,Q)$
\beqa
Q^2&:\quad&\frac{8}{3} S_1(M) \left(\HS_1^4-\HS_1 \HS_3\right)
- 2 \HS_1 \left(2 \HS_{3,1}+\HS_4\right)
- 2 \left(5 \HS_{3,2}-\HS_{2,3}+6 \HS_{4,1}-12 \HS_{3,1,1}+\HS_5\right)\nonumber\\&&
-\frac{16}{3}  \HS_1^5
+2 \HS_2 \HS_1^3
+\frac{22}{3}  \HS_3 \HS_1^2
\eeqa
does not contain of the special sums, while the result for the third power of $Q$
\beqa
Q^3&:\quad&
-\frac{4}{3} S_1(M) \HS_1 \left(6 \HS_{2,2}+\HS_1^4-4 \HS_3 \HS_1-3 \HS_4\right)
+2 \HS_1^2 \left(8 \HS_{2,2}-2 \HS_{3,1}+\HS_4\right)\nonumber\\&&
+64 \HS_1
\left(
S_{-2,1,-2}(M)
+S_{-2,1,2}(M)
+S_{2,1,-2}(M)
+S_{2,1,2}(M)
\right)\nonumber\\&&
+\frac{8}{3} \HS_1
\left(
-5 \HS_{2,3}
+4 \HS_{3,2}
+3 \HS_{4,1}
-6 \HS_{2,1,2}
+6 \HS_{2,2,1}
-12 \HS_{3,1,1}
+5 \HS_5\right)\nonumber\\&&
-15 \HS_{2,4}
+20 \HS_{3,3}
+9 \HS_{4,2}
+12 \HS_{5,1}
+6 \HS_{2,2,2}
-12 \HS_{2,3,1}
-36 \HS_{3,1,2}
-36 \HS_{3,2,1}\nonumber\\&&
-48 \HS_{4,1,1}
+24 \HS_{2,2,1,1}
+72 \HS_{3,1,1,1}
+\frac{8 \HS_1^6}{3}
-\frac{5}{3} \HS_2 \HS_1^4
-\frac{32}{3} \HS_3 \HS_1^3
+5 \HS_6
\eeqa
contains the special sums again.
This observation and its generalization to the higher special sums allows us to assume the minimal set of the most simple special sums,
which can enter into the basis for the rational part of $\reciT$.
So, the minimal basis for the rational part of $\reciT$ has 83 reciprocity-respecting sums of the transcendentality level 9  with argument $M/2$
\beqa
&&\{\HSB_9,\HSB_{1,2,6},\HSB_{1,3,5},\HSB_{1,4,4},\HSB_{1,5,3},\HSB_{1,6,2},\HSB_{1,7,1},\HSB_{2,1,6},
\HSB_{2,2,5},\HSB_{2,3,4},\HSB_{2,4,3},\HSB_{2,5,2},\HSB_{2,6,1},\nonumber\\&&
\HSB_{3,1,5},
\HSB_{3,2,4},\HSB_{3,3,3},\HSB_{3,4,2},\HSB_{3,5,1},\HSB_{4,1,4},\HSB_{4,2,3},\HSB_{4,3,2},
\HSB_{4,4,1},\HSB_{5,1,3},\HSB_{5,2,2},\HSB_{5,3,1},\HSB_{6,1,2},\nonumber\\&&
\HSB_{6,2,1},\HSB_{7,1,1},
\HSB_{1,2,1,1,4},\HSB_{1,2,1,2,3},\HSB_{1,2,1,3,2},\HSB_{1,2,1,4,1},\HSB_{1,2,2,1,3},
\HSB_{1,2,2,2,2},\HSB_{1,2,2,3,1},\HSB_{1,2,3,1,2},\nonumber\\&&
\HSB_{1,2,3,2,1},\HSB_{1,2,4,1,1},
\HSB_{1,3,1,1,3},\HSB_{1,3,1,2,2},\HSB_{1,3,1,3,1},\HSB_{1,3,2,1,2},\HSB_{1,3,2,2,1},
\HSB_{1,3,3,1,1},\HSB_{1,4,1,1,2},\nonumber\\&&
\HSB_{1,4,1,2,1},\HSB_{1,4,2,1,1},\HSB_{1,5,1,1,1},
\HSB_{2,1,1,1,4},\HSB_{2,1,1,2,3},\HSB_{2,1,1,3,2},\HSB_{2,1,1,4,1},\HSB_{2,1,2,1,3},
\HSB_{2,1,2,2,2},\nonumber\\&&
\HSB_{2,1,2,3,1},\HSB_{2,1,3,1,2},\HSB_{2,1,3,2,1},\HSB_{2,1,4,1,1},
\HSB_{2,2,1,1,3},\HSB_{2,2,1,2,2},\HSB_{2,2,1,3,1},\HSB_{2,2,2,1,2},\HSB_{2,2,2,2,1},\nonumber\\&&
\HSB_{2,2,3,1,1},\HSB_{2,3,1,1,2},\HSB_{2,3,1,2,1},\HSB_{2,3,2,1,1},\HSB_{2,4,1,1,1},
\HSB_{3,1,1,1,3},\HSB_{3,1,1,2,2},\HSB_{3,1,1,3,1},\HSB_{3,1,2,1,2},\nonumber\\&&
\HSB_{3,1,2,2,1},
\HSB_{3,1,3,1,1},\HSB_{3,2,1,1,2},\HSB_{3,2,1,2,1},\HSB_{3,2,2,1,1},\HSB_{3,3,1,1,1},
\HSB_{4,1,1,1,2},\HSB_{4,1,1,2,1},\HSB_{4,1,2,1,1},\nonumber\\&&
\HSB_{4,2,1,1,1},\HSB_{5,1,1,1,1}\}\label{RRSums9}
\eeqa
and two special sums with argument $M$
\beqa
\MHSB_{2,2,\underline{\bf 1},3}(M)&=&
\HSB_{2,2,1,3}(M)+\HSB_{-2,2,1,3}(M)+\HSB_{2,-2,1,3}(M)+\HSB_{2,2,1,-3}(M)+\HSB_{-2,-2,1,3}(M)\nonumber\\&&
+\HSB_{2,-2,1,-3}(M)+\HSB_{-2,2,1,-3}(M)+\HSB_{-2,-2,1,-3}(M)\,,\label{SpecSums9A}\\
\MHSB_{2,\underline{\bf 1},2,3}(M)&=&
\HSB_{2,1,2,3}(M)+\HSB_{-2,1,2,3}(M)+\HSB_{2,1,-2,3}(M)+\HSB_{2,1,2,-3}(M)+\HSB_{-2,1,-2,3}(M)\nonumber\\&&
+\HSB_{2,1,-2,-3}(M)+\HSB_{-2,1,2,-3}(M)+\HSB_{-2,1,-2,-3}(M)\,.\label{SpecSums9B}
\eeqa

\end{document}